\documentclass[twocolumn]{aastex61}

\shorttitle{Oscillations in Faculae during Their Lifetime}
\shortauthors{A.~Chelpanov, N.~Kobanov}
\begin{document}
\title{Oscillation Dynamics in Short-Lived Facula Regions during Their Lifetime}
\correspondingauthor{A.~\surname{Chelpanov}}
\email{chelpanov@iszf.irk.ru}
\author{Andrei~\surname{Chelpanov}}
\affil{Institute of Solar-Terrestrial Physics
                     of Siberian Branch of Russian Academy of Sciences, Irkutsk, Russia}
\author{Nikolai~\surname{Kobanov}}
\affiliation{Institute of Solar-Terrestrial Physics
                     of Siberian Branch of Russian Academy of Sciences, Irkutsk, Russia}

\begin{abstract}

We performed a multi-wave study of the oscillation dynamic in short-lived facula regions during their lifetime.
We studied oscillations in five regions, three of which belonged to the beginning of the current solar activity cycle, and two of them existed at the end of the previous cycle.
We found that in the facula regions of the current cycle, low-frequency (1\,--\,2\,mHz) oscillation dominated at the early stages of the faculae formation, while in the regions of the previous cycle, five-minute oscillations dominated at this stage.
At the maximal development phase of all the facula regions, the locations of observed low frequencies are closely related to those of the coronal loops.
These results support the version that the sources of the low-frequency oscillations in loops lie loops' foot points.
 
\end{abstract}

\section{Introduction} \label{sec:intro}

Solar faculae are small bright spots seen in white-light images, which correspond to bright magnetic points in the photosphere.
A region of faculae is usually cospatial with a plage, a bright region in the chromospheric images.
Faculae regions are well identified in small active regions or new/old active regions without sunspots.
Due to the prevalence and relatively large area, a facula region can play an important role in the processes of the energy exchange and transformation.
In addition, along with active regions containing sunspots, facula regions may be sources of solar flares.
Of all the manifestations of solar activity, solar flares have the most pronounced effects on the Earth, near-earth space, and the configuration of the magnetosphere.
Oscillations in facula regions have been studied for over half a century \citep{1965ApJ...141.1131O, 1967SoPh....2....3H}.
Most of the research on this topic has been done based on velocity signal oscillations.
It was noted that, in general, oscillations in facular regions are weakened in comparison with the surrounding quiet Sun.
Five-minute oscillations stand out among the registered oscillations both in the photosphere and in the chromosphere \citep{2006ASPC..358..465C, 2007A&A...461L...1V}.
It was also found that the observed oscillations are non-uniformly distributed over the area of facula regions: three- and five-minute oscillations prevail within the chromospheric cells, and oscillations of lower frequencies 1,2\,--\,2,0\,mHz prevail over the network \citep{2011SoPh..268..329K}.
Mainly three-minute waves propagate upwards in the central parts of facula regions in the chromosphere, and five-minute waves propagate at the periphery \citep{2009ApJ...702L.168D}.
Oscillations with long periods, up to 4.5 hours, were also studied in facular regions \citep{2016Ge&Ae..56.1052S, 2017A&A...598L...2K}.

A small developing facula region soon after appearing on the surface on the one hand may be similar to a sunspot, because, as a sunspot, it is a case of emergence of a compact region of an enhanced magnetic field on the surface. Magnetic field configuration in these regions, however, is more chaotic than in sunspots \citep{1997ApJ...474..810M}, which complicates their study.
This is especially pronounced in the chromosphere, where the magnetic pressure becomes comparable to the plasma pressure, and the environment becomes much less homogeneous.
\citet{2015ARep...59..968C} found that 3\,--\,6\,mHz oscillations dominate in the chromosphere over the magnetic field concentration nodes, while the dominant frequencies over the peripheral regions are lower -- 1.5\,--\,3\,mHz.
The characteristics of the magnetic field oscillations in the magnetic hills of facula regions -- local maxima of the magnetic field strength -- were studied \citep{2005A&A...437.1055M,2007SoPh..246..273K,2015ARep...59..968C,2016SoPh..291.3329C,2021arXiv210111998J}.

\citet{2019Ap&SS.364...29S} studied oscillations in the magnetic nodes of facula regions and revealed a wide range of frequencies in them, which are influenced by changing physical parameters in the nodes.

In the works listed above, the characteristics of oscillations were investigated in the fully-formed facula regions in the phase of their maximum development.
A facula region, however, is a dynamic object, and studying oscillation characteristics during its evolution is of considerable interest.
As short-lived facula regions we define regions, whose full life cycle can be observed on the visible hemisphere of the Sun under 14 days.
In order to distinguish them from other short-lived objects as, e.g., bright coronal points (CBPs), we chose the lower life time limit to be three days, given the life time of 95\,\% of CBP does not exceed 20~hours \citep{2015ApJ...807..175A}.
Despite a considerable number of facula regions, few of them meet this criterion.
Earlier, \citet{2020SoPh..295...94C} observed the region that was registered as the first active region of the current 25th cycle.
This work describes the dynamics of the oscillation spectra at different stages of the active region development.
Two questions arose based on these results: first, will the obtained results be typical of the other facula regions in this cycle;
second, are there differences in the oscillation dynamics of active regions of the ending previous cycle and the beginning of the new one?
Of the four facula regions that are added to the analysis, two regions belong to the current cycle, and two regions belong to the ending of the previous cycle.

The aim of this work is to try to answer this questions and to expand our knowledge of the oscillatory processes in the short-lived facula regions.

\section{Data and Analysis}

For this paper, we used the Solar Dynamics Observatory (SDO) data for five facula regions that appeared and developed at the visible side of the Sun.
In the analysis we used the Atmospheric Imaging Assembly (AIA) 1600\,\AA, 304\,\AA, and 171\,\AA\ channels that represent the chromosphere, transition region, and lower corona, respectively.
The instrument provides full-disk observations in the emission intensity.

The cadence of these data is 12\,s (24\,s for the 1600\,\AA\ channel); the spatial sampling is 0.6 arcsec per pixel.
To prepare and de-rotate the SDO data, we used the algorithms provided by the sunpy core package \citep{sunpy_community2020}.
To construct distributions of oscillation frequencies over the active regions’ areas, we used the Fast Fourier Transform algorithm: for each spatial point, we calculated the oscillation spectra, then, based on these spectra, we found the central frequency of the one\,mHz-wide window that yielded the highest integral oscillation power.

The five facula regions that we analyse appeared on the surface of the Sun on 13 February 2019, 09 December 2019, 6 July 2019, 25 September 2020, and 23 October 2020.
They evolved into fully-formed medium-sized regions and mostly decayed while still on visible side of the Sun.
The oscillation dynamics of the facula region that appeared on 6 July 2019 was considered in \citet{2020SoPh..295...94C}.

Spectral observations of the facula region that appeared on 6 July 2019 were performed with the use of the ground-based Horizontal Solar Telescope of the Sayan Solar Observatory.
The observations were carried out on 7, 8, 10, and 11 July while the region moved across the visible side of the Sun.
We used spectrograms of two chromospheric lines: H$\alpha$ 6563\,\AA\ and He\,\textsc{i} 10830\,\AA.
The temporal rate of the series is 1.5\,s, and the spectral sampling is around 16\,m\AA\.
We estimate the real spatial resolution to be 1.0\,--\,1.5\,arcsec due to the atmospheric limitations.
Based on the spectrograms, we calculated intensity and line-of-sight velocity signals.

We divided the lifetime of the facula regions into four phases: i) emergence -- the appearance of increased magnetic field at the photospheric level and bright structures in the chromospheric channels; ii) growth -- increase in the size and brightness of a region; iii) maximal phase -- the size and brightness are at their peaks, and the coronal loop system in the 171\,\AA\ channel has been fully developed; iv) decay -- decrease in brightness and a fading coronal loop system.
Based on the FFT spectra, we calculated the frequencies that dominated within the facula region borders at each phase of the region lifetime.
The region borders were determined based on the 1600\,\AA\ channel images. Table~1 shows the obtained frequency ranges divided into 0.5\,mHz intervals.
For acula Region~2 these intervals are 1\,--\,2\,mHz.

\section{Results}

\begin{figure}
\centerline{
\includegraphics[width=7cm]{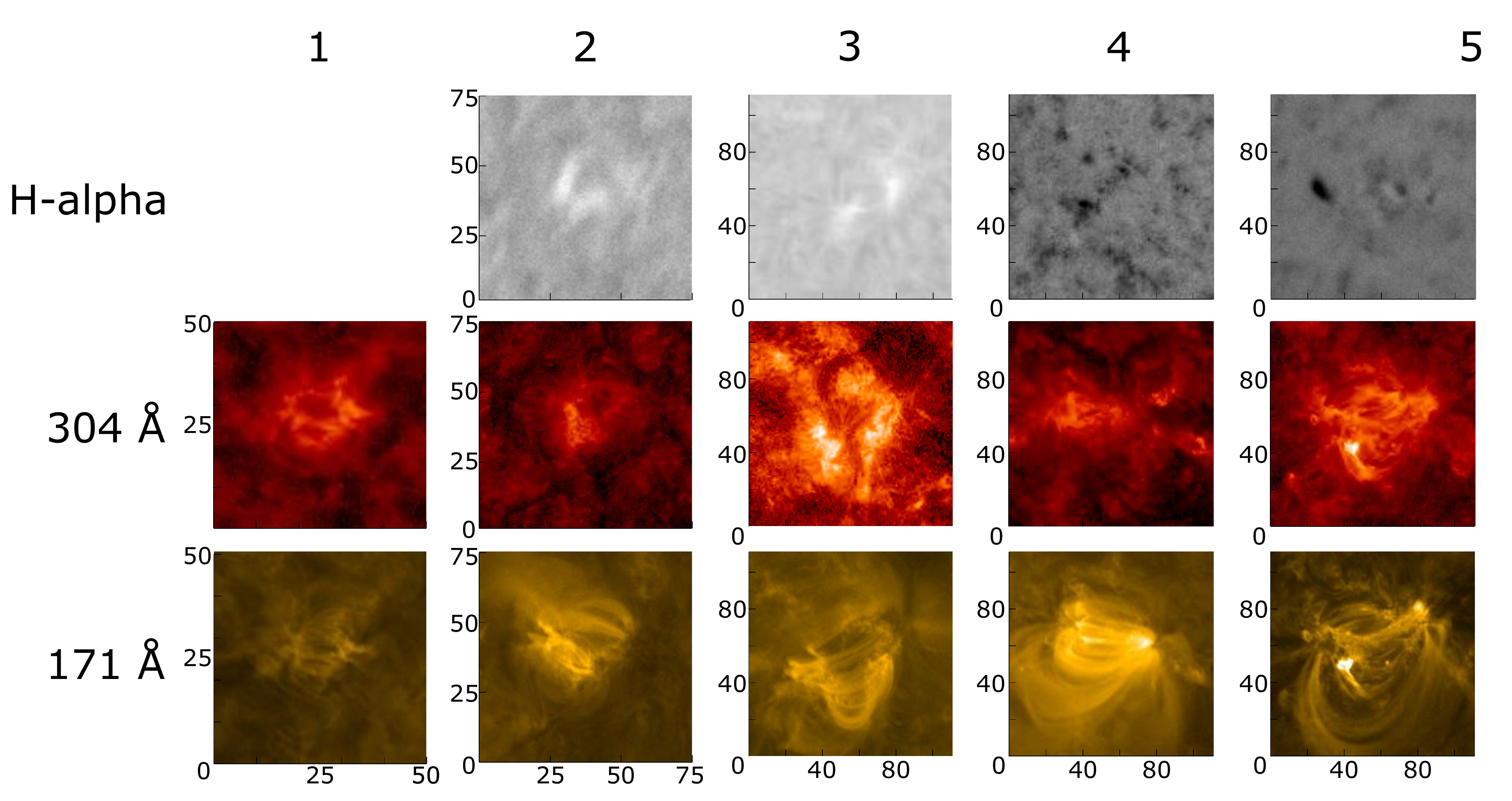}
}
\caption{Images of the five analysed facula regions at their maximal phases in the Big Bear Solar Observatory H$\alpha$ data and in the 304\AA\ and 171\AA\ AIA channels. The H$\alpha$ data are missing for the dates of Facula region~1.}
\label{fig:bands}
\end{figure}

\begin{figure}
\centerline{
\includegraphics[width=10cm]{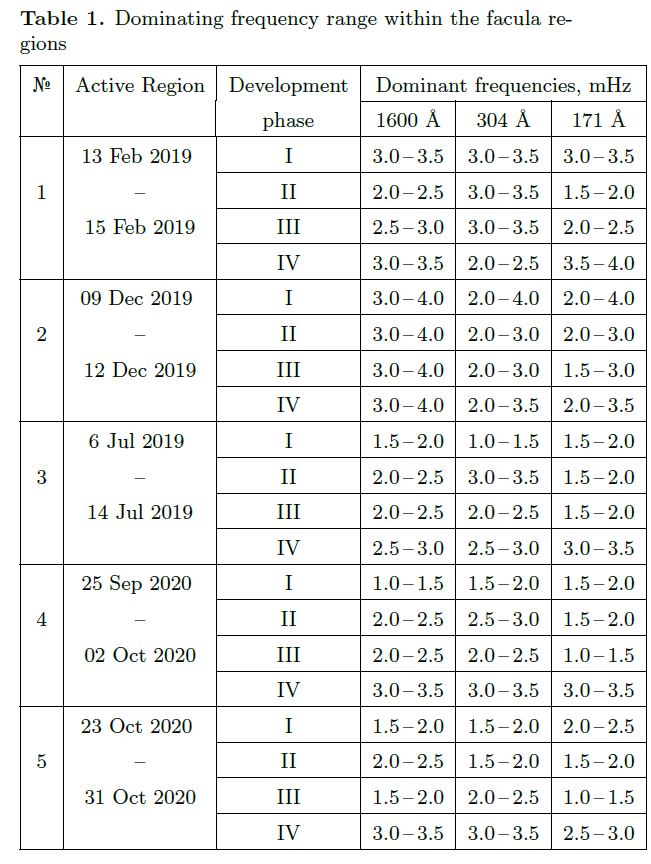}
}
\caption{Table~1.}
\label{fig:table}
\end{figure}

The analysis of the spatial distributions of oscillations in facula regions is more complicated than in sunspots because of a more complex field topology: a facula region may contain several nodes of magnetic field concentration, often of different polarities.
In general, the shapes of facula regions are more chaotic and less symmetrical than those of sunspots.
Figure~\ref{fig:bands} shows the morphological features of the studied regions at their maximal phase in the chromosphere, transition region, and lower corona (the H$\alpha$ line, the 304\AA\ and 171\AA\ AIA channels).
The information on the strength and inclination of the magnetic field in faculae is diverse and somewhat contradictory.

The three facula regions of the new solar activity cycle show similar oscillation dynamics.
At the emergence and growth stages, low frequencies prevail in the 1600\,\AA\ and 304\,\AA\ signals.
In the 171\,\AA\ channel signals, low frequencies dominate at the first three stages, especially at the maximal phase.
At the decay phase, periods close to 5 minutes dominate in all the observed AIA channels.

Table~1 shows that oscillation spectra in the facula regions of the previous activity cycle differ from those of the current cycle.
At the emergence phase, 3.0\,--\,3.5\,mHz oscillations dominate in the observed AIA channels in the regions of Cycle~24, while for the regions of the current cycle, the dominant range is 1.5\,--\,2.0\,mHz at this phase.
In Facula Region~1, five-minute oscillations (3.0\,--\,3.5\,mHz) dominate in the 304\,\AA\ signals up to the decay phase, when the frequency drops to 2.0\,--\,2.5\,mHz.
This is probably connected with the submergence of the coronal loop system in this active region during this phase.
At the decay phase of the facula regions of Cycle~24, nearly all the channels show oscillations with periods close to 5 minutes, as well as in the regions of the current cycle.

\begin{figure}
\centerline{
\includegraphics[width=9cm]{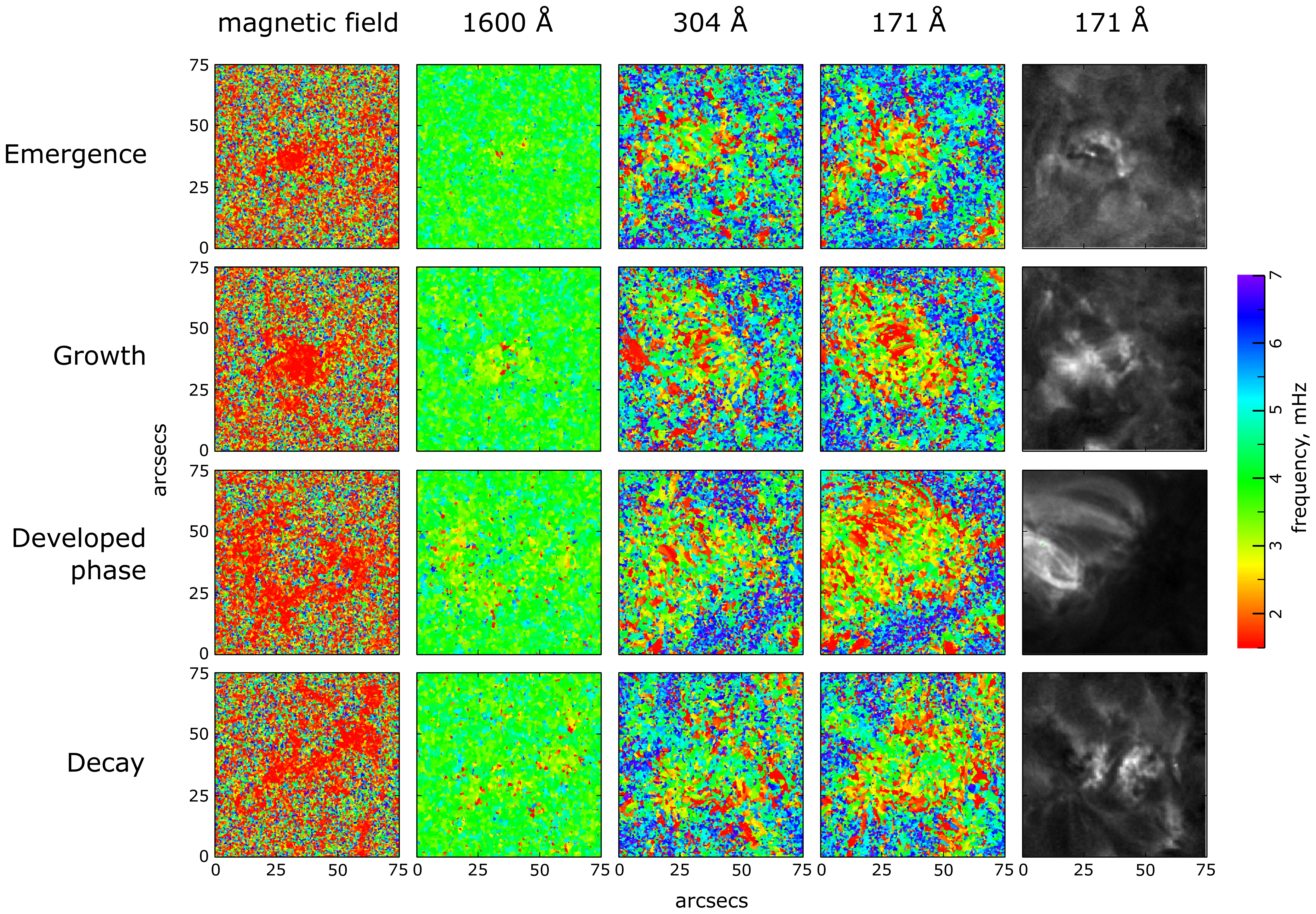}
}
\caption{Spatial distribution of the dominant oscillation frequencies and coronal loops in the 171\,\AA\ channel during four development phases in Facula Region~2 in Table~1.}
\label{fig:NewestF}
\end{figure}

\begin{figure}
\centerline{
\includegraphics[width=9cm]{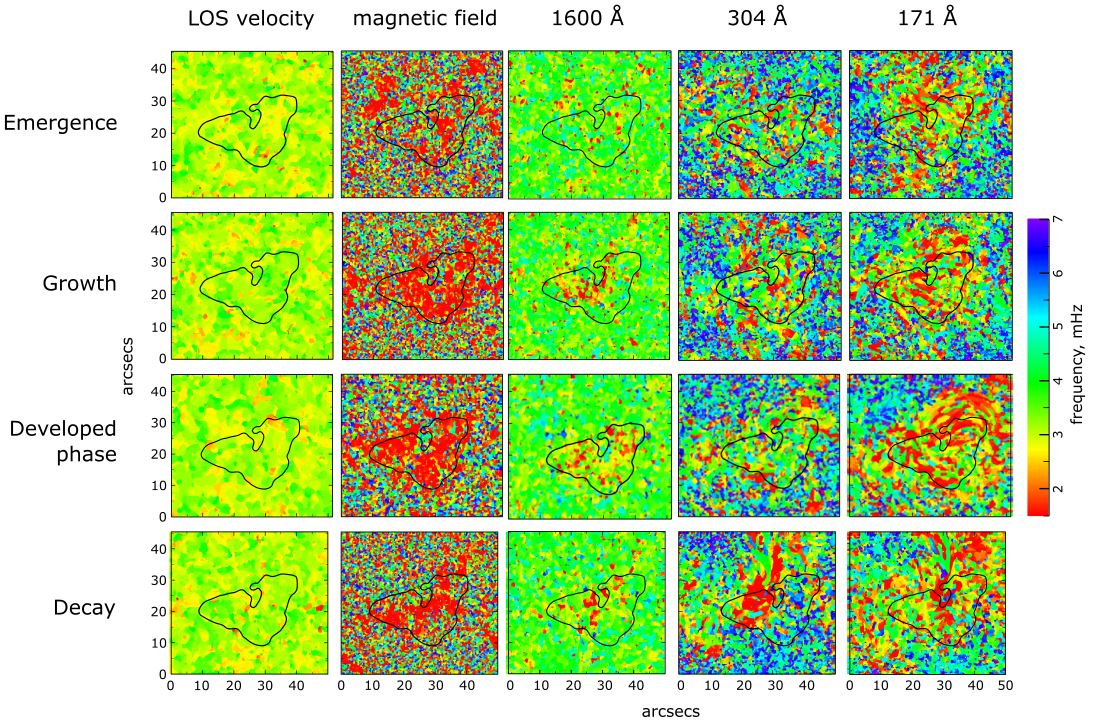}
}
\caption{Spatial distribution of the dominant oscillation frequencies during four development phases in Facula Region~1 in Table~1. The contour shows the boundaries of the region in its fully developed phase as seen in the 1600\,\AA\ channel.}
\label{fig:0}
\end{figure}

More detailed information on the spatial frequency distribution in the facula regions of Cycle~24 during the evolution can be found in Figures~\ref{fig:NewestF},~\ref{fig:0}.
In Figure~\ref{fig:0} with the three spectral channel used in Table~1, we show the spatial distributions of the dominant frequencies in the photospheric LOS velocity and magnetic field HMI signals.
The spatial frequency distribution of the photospheric LOS velocity signals provide little information.
Mostly five-minute oscillations prevail at the photospheric level at all the evolution phases of this facula region in the LOS-velocity signals.
In the magnetic field signals, low frequencies dominate at the growth and maximal phases in the two facula regions of Cycle~24 (Figures~\ref{fig:NewestF}, \ref{fig:0}).
Note that for the first facula of the current cycle \citep{2020SoPh..295...94C}, the densest low frequency concentration in the magnetic field distribution was observed during the emergence phase.

\begin{figure}
\centerline{
\includegraphics[width=9cm]{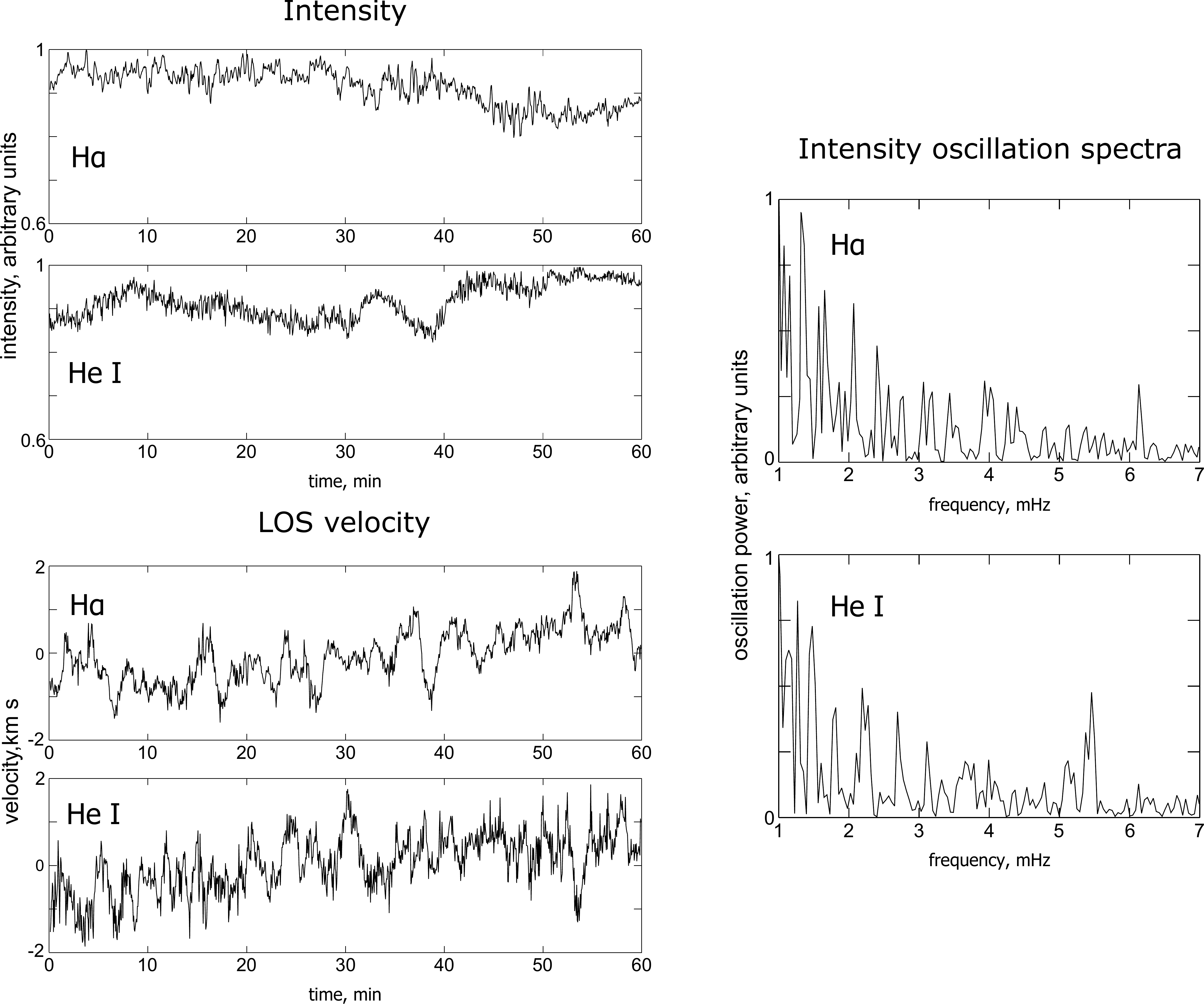}
}
\caption{Examples of intensity and LOS velocity signals and intensity oscillation spectra in the ground-based observations for the maximal phase of Facula Region~3.}
\label{fig:Signals-Spectra}
\end{figure}

\begin{figure}
\centerline{
\includegraphics[width=9cm]{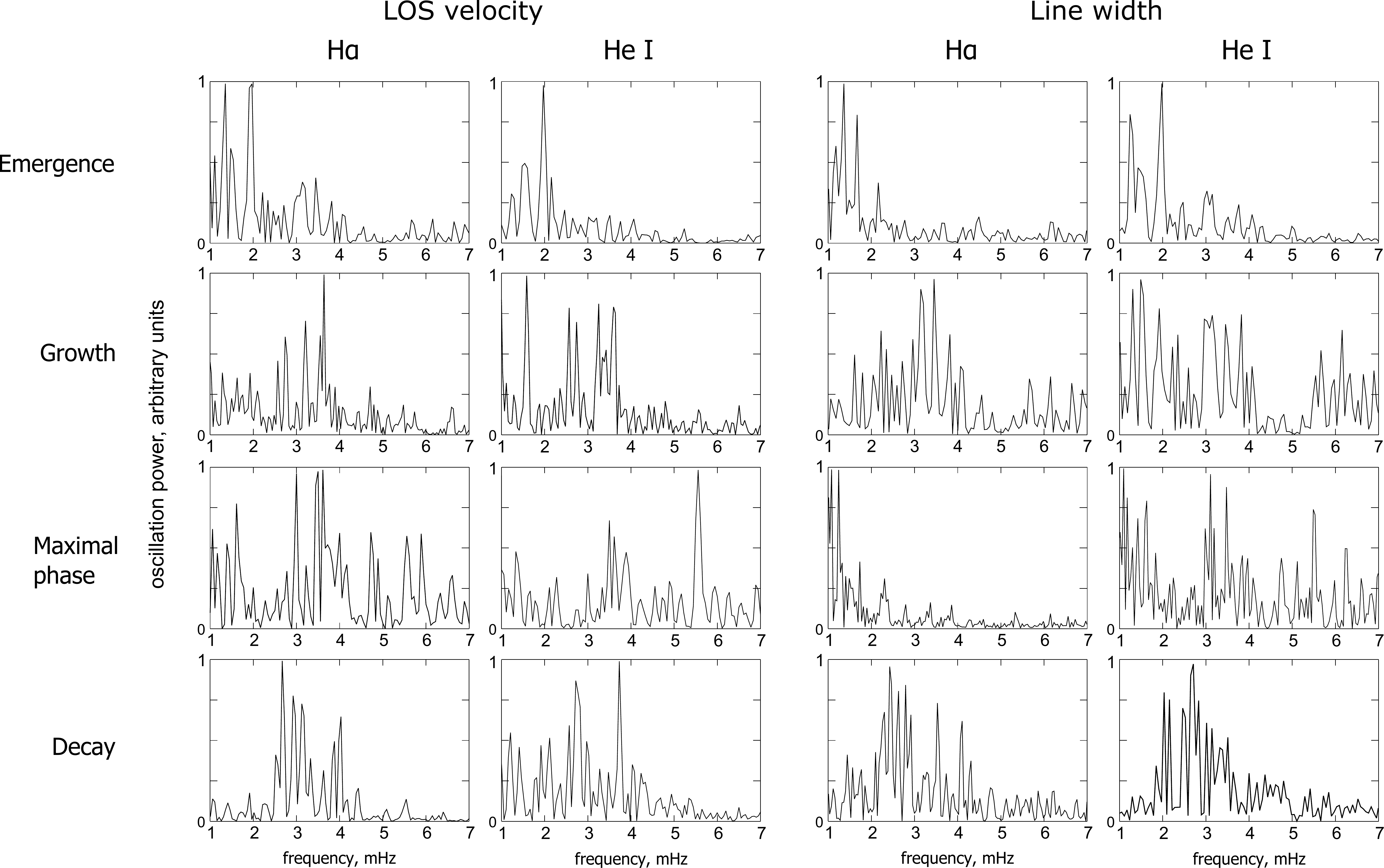}
}
\caption{Oscillation spectra of the chromospheric LOS velocity and line width signals at different phases of the development for Facula Region~3 in Table~1.}
\label{fig:1}
\end{figure}

Unfortunately, spectral observations in the chromospheric H$\alpha$ and He\,\textsc{i} 10830\,\AA\ lines were only made for AR~12744 (Facula Region~3) in the interval July 7 to 11, 2019.
Nevertheless, it is useful to examine the spectra dynamics of the chromosphere oscillations in this region.
Figure~\ref{fig:Signals-Spectra} shows examples of raw intensity and velocity signals, and intensity oscillation spectra in two chromospheric lines.
The intensity signals contain more noise as compared to the velocity signals, so below we focus on the LOS velocity oscillation spectra.
In the chromosphere, during the time from the emergence of the region to the development into a fully-formed facula region, the spectral composition of the velocity and line-width oscillations changes significantly.
From spectral observations of the velocity in the region that appeared on 6 July 2019, it can be seen that low frequencies dominate in the facular chromosphere at the first phase, and starting from the growth phase, frequencies in the range of 3-4 MHz become dominant, i.e., five-minute oscillations (Figure~\ref{fig:1}).
The high-frequency peaks that appeared in the spectra of the maximal phase, especially in the He\,\textsc{i} 10830\,\AA\ line, may be explained by the reduction in the relative weight of the low frequencies in the chromosphere at this phase, as well as by the complicated He\,\textsc{i} line formation mechanism, which involves UV radiation from the upper layers of the atmosphere.
We should note that the presence of low frequencies in the intensity and half-width signals is typical of the maximal phase (Figure~\ref{fig:1}).

\section{Discussion}

Usually, developed facula regions show the presence of strong five-minute oscillations in the photosphere and chromosphere, but the stages of appearance and growth show a concentration of low frequencies of 1–-2\,mHz (Figure~\ref{fig:0}) in the magnetic field signals.
They may occupy a significant part of the area of the emerging active region \citep{2020SoPh..295...94C}.
Low frequency areas are also found in the transition region \citep{2011SoPh..268..329K, 2015SoPh..290..363K}.
Such a spatial distribution of the dominant oscillation frequencies in small facula regions fundamentally differs from that observed in fully-formed sunspots.
In sunspots, the frequencies in the center, on the contrary, are high, and they decrease with the distance from the center.
Low frequencies of 1–-2\,mHz are observed only at the very edges of the spot, or even beyond them \citep{2004ARep...48..954K,2013SoPh..288...73M,2015SoPh..290..363K}.

At the earliest stage, the active facular region does not manifest itself prominently in the corona, and in the distribution of dominant frequencies it weakly stands out against the surrounding background.
Later, at the stage when an active region shapes into a more developed structure, a system of coronal loops is formed above it.
At this time, the elongated shapes of coronal loops appear in low frequencies (1\,--\,2\,mHz) in the distribution of the dominant frequencies of the active region in the 171\,\AA\ coronal line (Figures~\ref{fig:NewestF},~\ref{fig:0}).

The distributions of the magnetic field signals show that, at the earliest phase of the region formation, patches of low frequencies appear at the location of the facula region. Such patches can be also seen in the 1600\,\AA\ emission, though in a less pronounced form. Later, in the magnetic field these frequencies occupy the most part of the facular area.

In the AIA channels spectra, we note some differences in the oscillation spectra dynamics between the facula regions of the current cycle and those of the end of the previous cycle.
At present, we cannot explain these differences.
Such a research would require a larger set of observational material.
Probably, the causes can be found in the magnetic field topology characteristics connected to the phases of the solar activity cycles.
One should bear in mind that the active region loop system comprises loops of different height \citep{1996SoPh..168..115M,2012ApJ...757..167K}, which influences the characteristics of the dominant frequencies distribution in different spectral channels.
We should note that the facula regions at the end of the previous cycle existed for much shorter time than the regions of the current cycle.
The maximal LOS magnetic field strength in one of them, however, was over 1\,kG, which is higher than in some other regions in the table.
Note also that the three facula regions of the current cycle in Table~1 located in the southern hemisphere, while the two regions of Cycle~24 located in the northern hemisphere.
The north-south asymmetry of the solar activity \citep{2015LRSP...12....1V,2019LRSP...16....2M} may play a role in the evolution characteristics of these regions.

We assume that the low frequencies observed in the chromospheric distributions and the low frequencies observed in the loops in the corona are interconnected and represent the same process.
The frequency distributions show that the low frequencies correspond to the location of the loops in the layers of the atmosphere.
At the initial stages of the facula region formation, magnetic tubes rising to the surface penetrate only the lower layers. At the same time, the lower layers of the facula region are filled with low frequencies.
At subsequent stages, magnetic structures rise to coronal heights and form coronal loops, where the distributions of low frequencies outline the pattern of these loops.

\begin{figure}
\centerline{
\includegraphics[width=9cm]{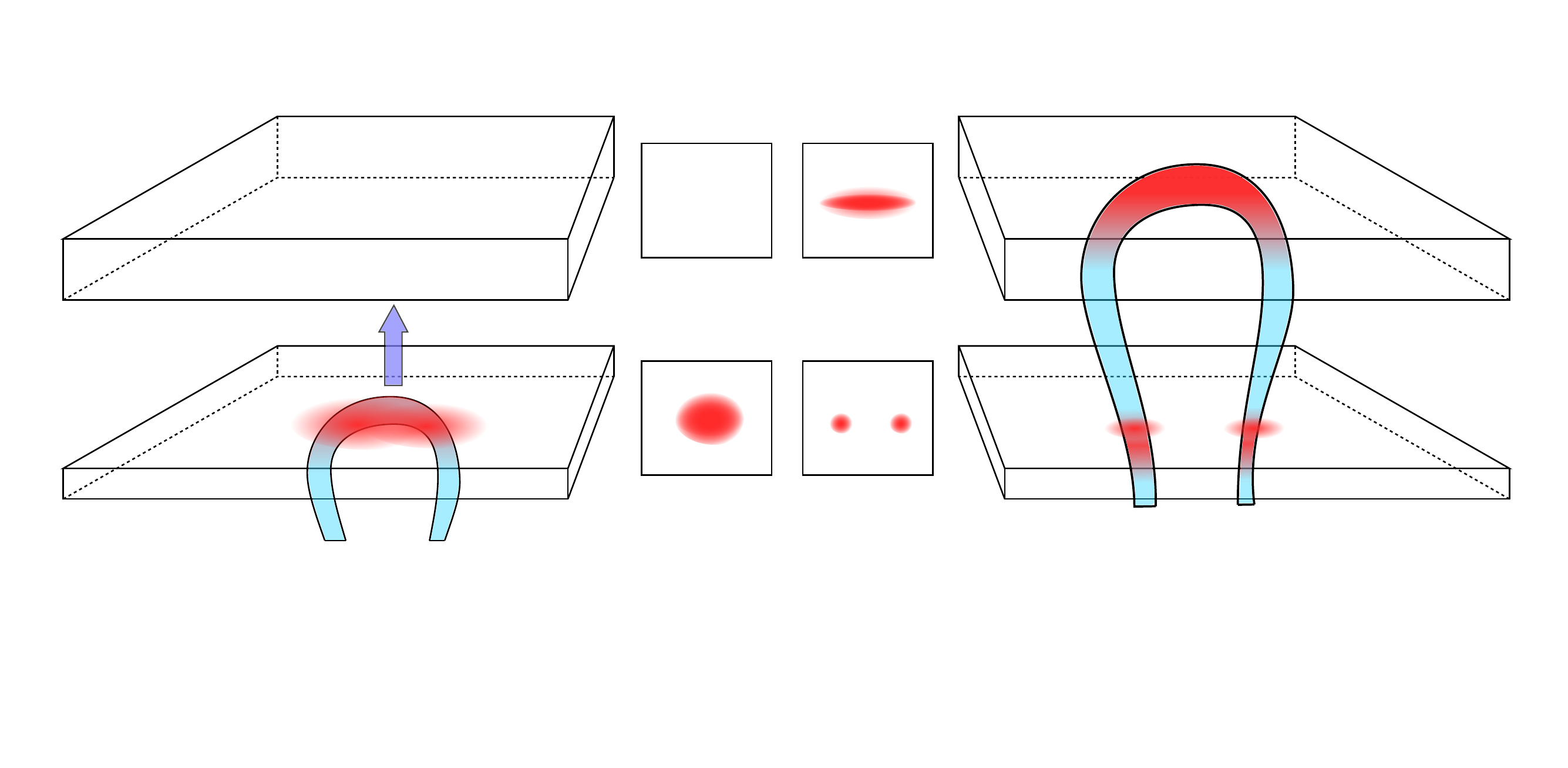}
}
\caption{Sketch showing the position of a coronal loop in a facula region relative to two layers and the locations of low frequencies in them.}
\label{fig:F-Sketch}
\end{figure}

Figure~\ref{fig:F-Sketch} shows a schematic representation of a magnetic loop at the time of a facula region emergence and at the developed stage.
On the left, the loop only reaches the lower atmosphere levels of the photosphere and chromosphere.
On the right, the loop's upper part is at the coronal heights.
The panels in the middle represent what the distributions of low frequencies look in the chromosphere and lower corona.
The red areas show the observed low frequency regions.
At the low layers, only vertical footpoints of the loops are observed, where they occupy smaller areas.
In this case, the pattern of low frequencies in the distribution of oscillations in the low layers breaks up into separate small domains.
At the last phase of the facula region evolution before its decay, the picture of the coronal loops dissipates. Low frequencies become sometimes more present at the lower level, in the transition region (for example, see Figure~\ref{fig:0}).
This may indicate that loops sink lower from the corona in the process of their disappearing.

Such a picture of the relation between low frequencies and magnetic coronal loops supports the presence of some kind of wave-guiding mechanism for the oscillations propagating from the foot points.
Earlier, \citet{2020SoPh..295...94C} showed that low frequencies in coronal loops indeed propagate as variations in brightness separately from their transverse oscillations as a whole.

\section{Conclusions}

In this work we studied the dynamics of the spatial distribution of oscillation frequencies in five facula regions, two of which belong to the end of Cycle~24, and the three others belong to the beginning of Cycle~25.
The conclusions are as follows:

\begin{enumerate}
  \item We established that the dynamics of the oscillation spectra is similar in different facula regions of the current cycle, while the regions of Cycle~24 show a number of differences.
  \item During the facula region life cycle, the spatial distribution of the dominant frequencies shows significant changes in the intensity and LOS magnetic field signals.
In the LOS velocity oscillations at the photospheric level, these changes are insignificant.
At the maximum phase, the spatial distribution of low frequencies in the 171\,\AA\ line corresponds to the locations of the coronal loop apexes.
At the decay phase, five-minute oscillations dominate in the intensity signals of all the AIA channels.
These characteristics are observed in all the studied faculae.
  \item For the facula regions of the current cycle, it was found that immediately before the first signs of brightening of the observed regions in the EUV lines, the signals of the magnetic field and the 1600\,\AA\ line intensity show concentrations of low-frequency oscillations (1\,--\,2\,mHz) in the central parts of the future facula region, while in the facula regions of the previous cycle, five-minute oscillations dominated at this phase.
  Besides, in one of the facula regions of Cycle~24, low frequencies (2.0\,--\,2.5\,mHz) in the 304\,\AA\ channel become more apparent only at the decay phase.
  This was probably caused by the submerging coronal loop system.
\end{enumerate}

We assume that in the facula regions of the current cycle, the observed spatial concentration of low-frequency oscillations in the signals of the magnetic field, the 1600\,\AA\ channel intensity, or the LOS velocity in the H$\alpha$ and He\,\textsc{i} 10830\,\AA\ chromospheric lines may be considered as one of the early predictors of the active region formation.

\acknowledgments

The study was funded by RFBR, project number 20-32-70076 and Project No.\,II.16.3.2 of ISTP SB RAS. Spectral  data were recorded at the Angara Multiaccess Center facilities at ISTP SB RAS. We acknowledge the NASA/SDO science team for providing the data.
We are grateful to an anonymous reviewer for suggestions that helped improve the text.

\bibliography{Chelpanov}

\end{document}